\documentclass[%
superscriptaddress,
Preprint,
Nopreprintnumbers,
 amsmath,amssymb,
 aps, 
]{revtex4-2}
\usepackage[version=4]{mhchem}
\usepackage[T1]{fontenc}
\usepackage{graphicx}
\usepackage[english]{babel}
\usepackage{dcolumn}
\usepackage{nicefrac}
\usepackage{booktabs}
\usepackage{bm}
\usepackage{hyperref}
\usepackage[mathlines]{lineno}
\linenumbers\relax 
\usepackage{siunitx}
\usepackage{float}
\newcommand{\ZGa}{$\text{Z}^{\text{ps}}_\text{Ga}$}
\newcommand{\ZN}{$\text{Z}^{\text{ps}}_\text{N}$}
\newcommand{\ZM}{$\text{Z}^{\text{ps}}_\text{Mt}$}
\begin{document}
\nolinenumbers
\preprint{APS/123-QED}

\title{Two-dimensional carrier gas at a polar interface without surface band gap states: A first principles perspective}

\author{Federico Brivio}
 \affiliation{Department of Molecular Chemistry and Materials Science, Weizmann Institute of Science, Rehovoth 76100, Israel \\
 }%
  \affiliation{Department of Electrical and Computer Engineering, Technion, Haifa, Israel}
 \email{federico.brivio@weizmann.ac.il}
  \author{Andrew M. Rappe}
 \affiliation{Department of Chemistry, University of Pennsylvania, Philadelphia, PA, 19104 USA}%
 \author{Leeor Kronik}
 \affiliation{Department of Molecular Chemistry and Materials Science, Weizmann Institute of Science, Rehovoth 76100, Israel \\
 }%
 \author{Dan Ritter}
  \affiliation{Department of Electrical and Computer Engineering, Technion, Haifa, Israel}
 \email{ritter@technion.ac.il}


\date{\today}

\vspace{2 cm}
\begin{abstract}
We present first principles calculations of the interface between GaN and strained AlN, using a slab model in which polarization is compensated via surface fractional-charge pseudo-hydrogen atoms. We show that an interface two-dimensional carrier electron or hole gas emerges naturally in response to different compensating surface charges, but that this need not involve in-gap surface states.   

\end{abstract}
\maketitle

\newpage
There has been a long-standing interest in the formation of a two-dimensional electron gas (2DEG) or hole gas (2DHG) in III-N  heterostructures~\cite{zgur1995high,khan1996microwave,binari1997aigan,gaska1997high,wu1997bias,meneghini2021gan,wellmann2022wide}.    
It is well understood that the 2D carrier gas (2DCG) partially compensates the interface polarization charge~\cite{asbeck_piezoelectric_1997,ambacher_two-dimensional_1999}. 
Early on, the origin of this charge has been attributed to the presence of ionized band-gap states at an external surface, often referred to as surface donors \cite{ibbetson2000polarization,miao2010oxidation,higashiwaki2010distribution}. Newer experimental studies, however, suggest that a 2DCG can be observed even when surface donors cannot fully explain its origin \cite{tapajna2017investigation,sayed2019net,ber2019measurement,liu2020improved,rrustemi2021investigation}. 
Understanding how this comes about calls for a first principles computational investigation of a polar interface, where surface states can be completely suppressed by construction.

Here, we provide such an investigation, based on an extension of a method recently suggested by Yoo \textit{et al}.\ \cite{yoo_finite-size_2021} for the cancellation of the macroscopic electric field in calculations of permanently polar slabs. 
We find that, indeed, surface band gap states are not required for generating free electrons or holes. 
Instead, these arise directly from the balance of surface and interface charges with the material polarization. 

\begin{figure}
\includegraphics[width=1\textwidth]{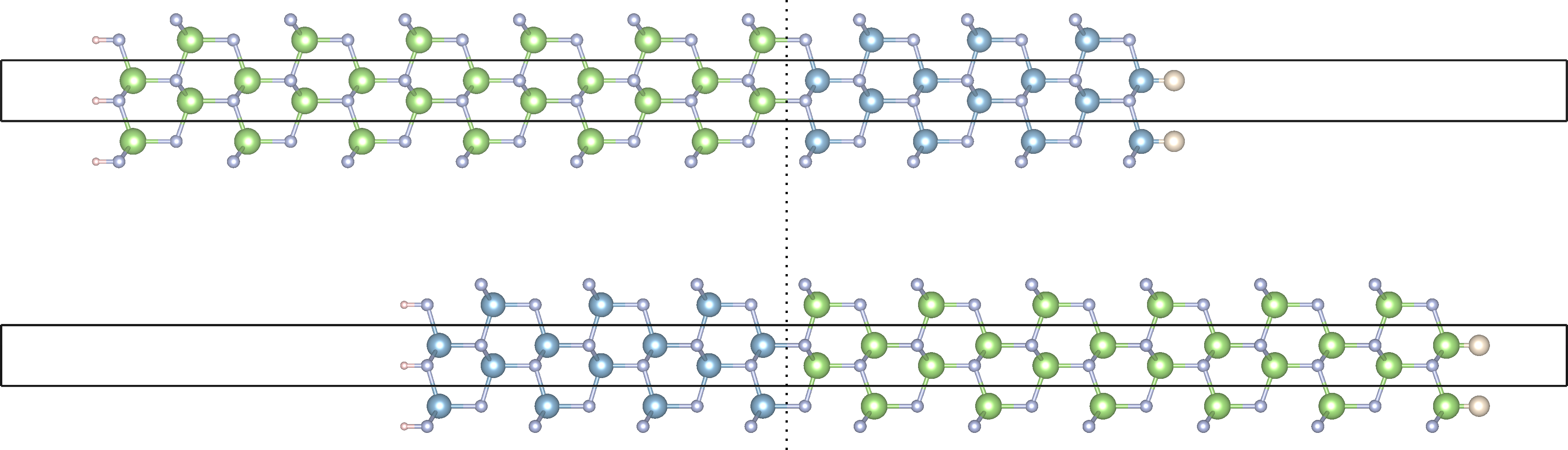} 
\caption{\label{fig:struct} Atomic structures of two different interfaces formed by joining a 12-layer GaN slab with a 7-layer strained AlN slab.
The black line indicates the slab unit cell. Ga atoms are green, Al atoms are light blue, and N atoms are small grey-blue spheres. Periodic images in the vertical direction are added for clarity. 
In the top (bottom) panel, the sAlN slab is attached to the Ga-face (N-face) of the GaN slab. The interface is placed at the center of the slab and is indicated by a dotted line. 
The external surfaces are passivated with two different pseudo hydrogen atoms (see text for details), such that the one with the larger valence is indicated by a larger sphere.}
\end{figure}

We consider, as a particular example, the (0001) interface between GaN and laterally-strained AlN (sAlN) that is lattice-matched to GaN. 
Both materials possess the wurtzite (wz) structure ($P6_3mc$, n. 186).
We analyze two types of interfaces embedded in a slab, as shown in Figure \ref{fig:struct}, one where the sAlN is attached to the Ga-face and one where it is attached to the N-face.   
From simple chemical considerations, and as shown explicitly below, these interfaces are expected to give rise to a 2DEG and a 2DHG, respectively.

Basic electrostatic considerations suggest that a sheet of polarization charge must arise at the interface between two materials of different net polarization. 
However, it is clear that if the macroscopic electric field is nonetheless zero across the entirety of the slab, then the polarization charge must be fully compensated by free carriers, which would form the 2DCG.
It is well known that III-V semiconductor unit-cell surfaces in general~\cite{shiraishi1990new,huang2005surface} and III-N  unit-cell surfaces in particular~\cite{wang2001electronic,kempisty2012ab} can be electronically passivated by means of fictitious, pseudo-hydrogen (psH) atoms with fractional atomic numbers of $Z$=0.75 and $Z$=1.25 for psH bonded to column V and column III surface atoms, respectively. 
More recently, Yoo \textit{et al}.\ \cite{yoo_finite-size_2021} have shown that the polarization charge at the free surface of a slab comprising of one material can be compensated by means of some alteration of the $Z$ value of the psH atoms from their above values.
In the case of GaN, for example, the valence of the psH attached to the nitrogen-side, \ZN, is chosen as the polarization of the material, $P$ (calculated with respect to the non-polar, centrosymmetric hexagonal phase \cite{dreyer_correct_2016}), times the unit-cell facet area, $A_{(0001)}$. The valence of the psH attached to the gallium side, \ZGa, is then correspondingly chosen as \ZGa $=2-$\ZN.
Importantly, any fractional unit-cell charge we model could correspond to a physically realizable system with a long-range surface reconstruction or a partial coverage of charged adsorbates, in a larger supercell.

To extend the above approach to the study of an embedded interface as in Fig.\ \ref{fig:struct}, we attached psH atoms to both free (external) surfaces and adjusted the valence of the psH atom on each side to be the same as that of a slab comprising only of this material, with the same surface termination.  
Because the polarization values of both materials are different, this immediately implies that the sum of the $Z^{\text{ps}}$ at the two free surfaces is no longer equal to 2. 
Importantly, this means that the calculations now include a total fractional number of electrons. However, the system remains overall neutral because each fractional psH atom is neutral. 
If the above sum of charge is larger (smaller) than 2, then the slab is supplied with excess electrons (holes) that can accumulate at the interface, form a 2DEG (2DHG), and compensate the interface polarization charge. 
If this compensation is complete, the entire two-material slab will be devoid of a macroscopic electric field and one can then can study the properties of the ensuing 2DCG.

To utilize the above idea, we first calculated the bulk polarization values of both GaN and sAlN, given in Table \ref{tab:alfas}, using both the slab method \cite{yoo_finite-size_2021} and the Berry phase method \cite{dreyer_correct_2016}. The values for $Z^{\text{ps}}$ obtained based on the two methods are essentially identical and we proceed with the slab values for consistency.
The slab size has been chosen so that it is the minimal one for which a negligible macroscopic field could be identified. Still, even a small variation in the psH valence can lead to the build-up of a residual macroscopic field. We therefore increase (decrease) \ZGa (\ZN) by 0.001 electron compared to the values reported in table \ref{tab:alfas}, which we found to reduce the field further, to below 1 meV/\AA~at the center of the GaN or AlN side of the slab, which represents the numerical precision obtained from retaining three digits in the surface electronic charge.

\begin{table}[]
\caption{\label{tab:alfas} Values for the psH valence in single material models, calculated using the slab method~\cite{yoo_finite-size_2021} and the Berry phase method~\cite{dreyer_correct_2016}. The valence of the PsH attached to N atoms, \ZN, is obtained by multiplying the value of the polarization by the area of the (0001) facet. The valence of the PsH attached to Ga or Al atoms, \ZM, is obtained as \ZM $= 2-$ \ZN . The value of the facet area is $A_{(0001)} = 8.953$~\AA $^3$ for both materials.}

\begin{tabular}{S[table-format=4.3]S[table-format=4.3]S[table-format=4.3]S[table-format=4.3]S[table-format=4.3]} 
\toprule 
    & \multicolumn{2}{c}{Slab} & \multicolumn{2}{c}{Berry Phase} \\ 
 {System}  & {\ZN [e]}           & {\ZM [e]}        & {\ZN [e]}            & {\ZM [e]}      \\\midrule
{GaN} & 0.733       & 1.267      & 0.731      & 1.269      \\
{sAlN} & 0.672       & 1.328      & 0.674       & 1.326      \\ \bottomrule
\end{tabular}
\end{table}

All first principles calculations were performed using density functional theory (DFT) based on the planewave-based code VASP \cite{kresse_ab_1993,kresse_efficiency_1996} (Vienna \textit{Ab-initio} Simulation Package), in which core electrons are treated using projector augmented waves. Throughout, we used the generalized gradient approximation (GGA) of Perdew, Burke, and Ernzerhof (PBE) \cite{perdew_generalized_1996} for the exchange-correlation functional, with a planewave cutoff of 600 eV and a k-grid mesh of 5$\times$5$\times$1. Because we are dealing with a possibly polar slab in a periodical unit cell, we employed the dipole correction scheme of Neugebauer and Scheffler~\cite{neugebauer_adsorbate-substrate_1992}.
We placed the psH atoms, with the values given in Table \ref{tab:alfas}, at 1.05~\AA~from the N and 1.50~\AA~from the metallic atom. 
Atoms in the slab were originally positioned based on an optimized bulk unit cell, preserving the \ce{Ga-N} and \ce{Al-N} bond lengths at each interface. The sAlN internal coordinate and unit cell were fully optimized with the constraint of fixing the $a$ and $b$ lattice parameters to those of the GaN unit cell.

\begin{figure}[h]
\includegraphics[width=0.49\textwidth]{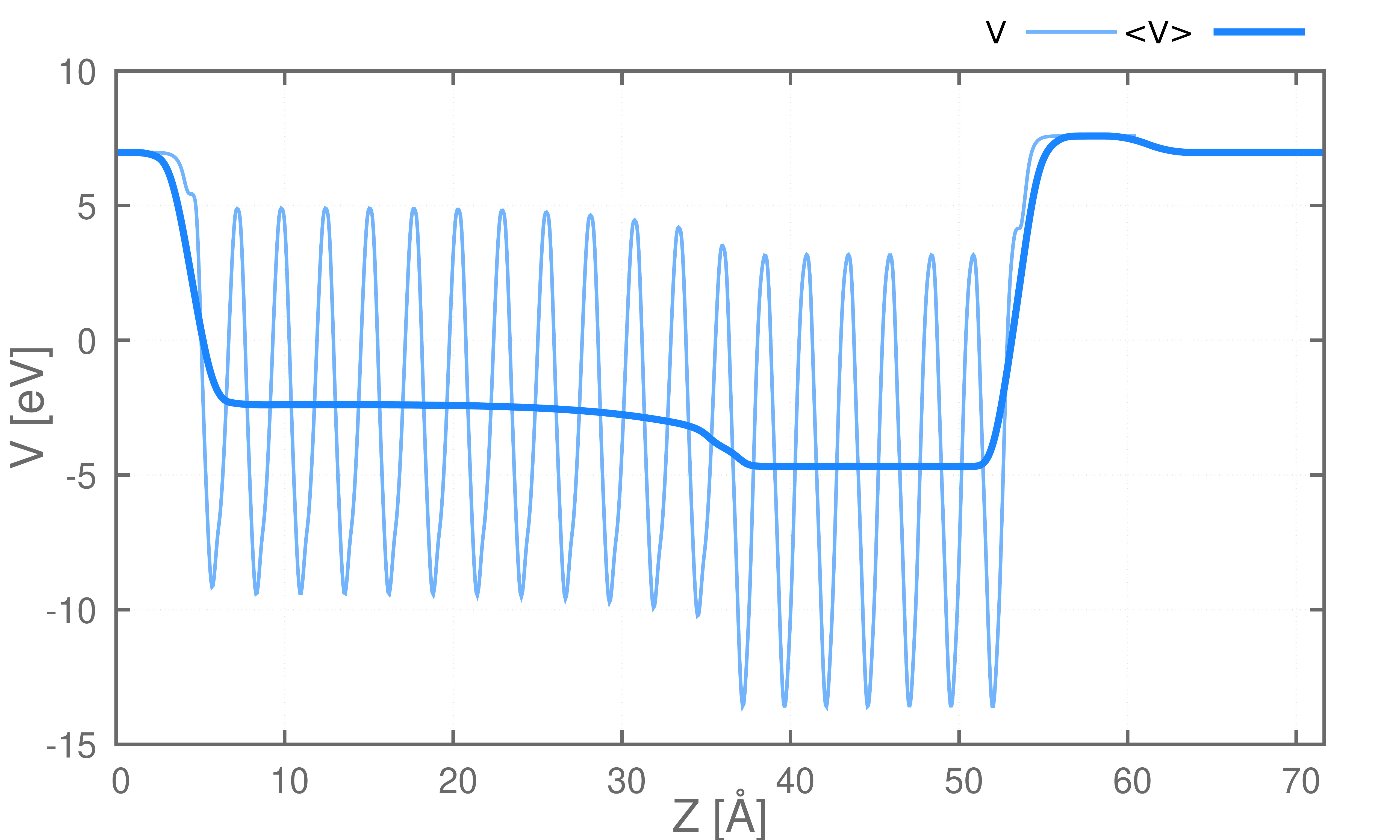} 
\includegraphics[width=0.49\textwidth]{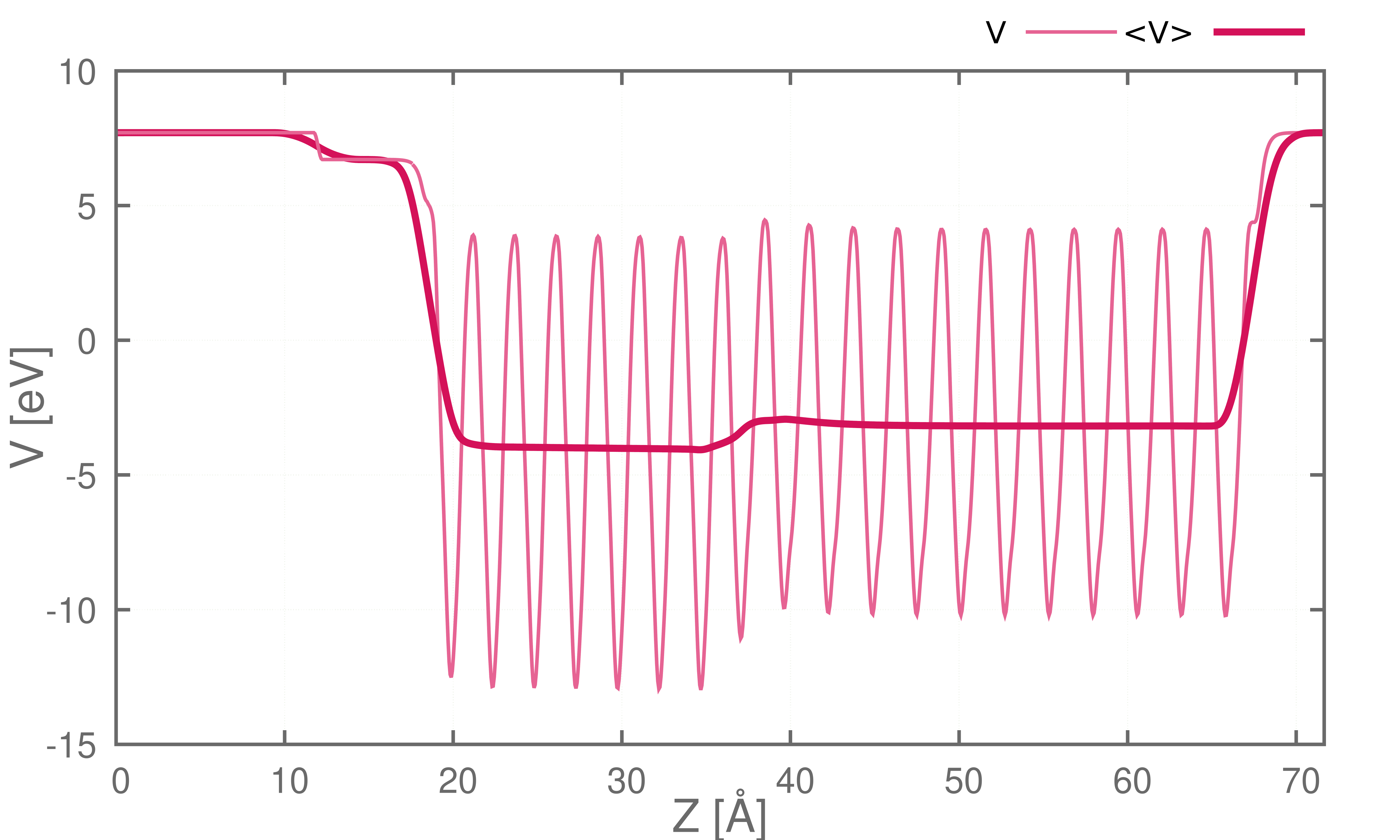}
\caption{\label{fig:potential} Local (thin line) and macroscopic (thick line) laterally averaged electrostatic potential across the slab structures of Fig.\ \ref{fig:struct}. The macroscopic field inside each material, far from a surface or interface, is close to zero ($<$ 1 meV/\AA), indicating a near-absence of polarization.}
\end{figure}

Figure \ref{fig:potential} shows the local electrostatic potential, obtained by combining the laterally-averaged ionic and Hartree potentials of the DFT calculation. It also shows the macroscopic electrostatic potential, obtained from the local one by averaging within a moving window along the z-direction, using the vaspkit \cite{wang_vaspkit_2021} tool. We applied the averaging twice, using the Ga-Ga and Al-Al interplanar distances as window widths.
The Figure shows that within a few atomic monolayers from the external surfaces and the interface, the field decays rapidly. Specifically, the macroscopic field evaluated at the center of the GaN or the AlN part of the slab, evaluated from the slope of the potential, is at most $\sim 1$ meV/\AA, indicating the near-absence of net polarization.

\begin{figure} 
\includegraphics[width=0.49\textwidth]{./FIG3Eb.png} 
\includegraphics[width=0.49\textwidth]{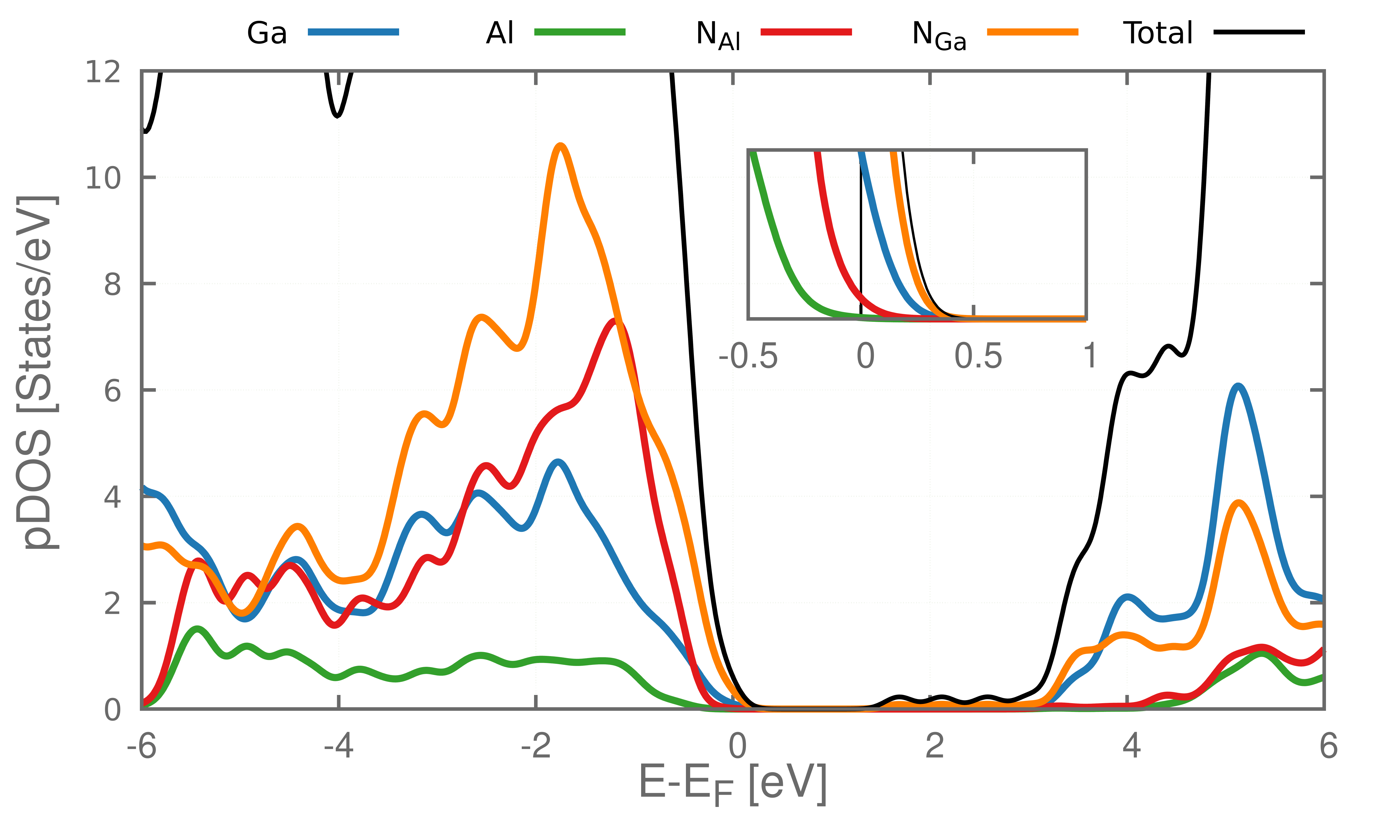}\\
\includegraphics[width=0.49\textwidth]{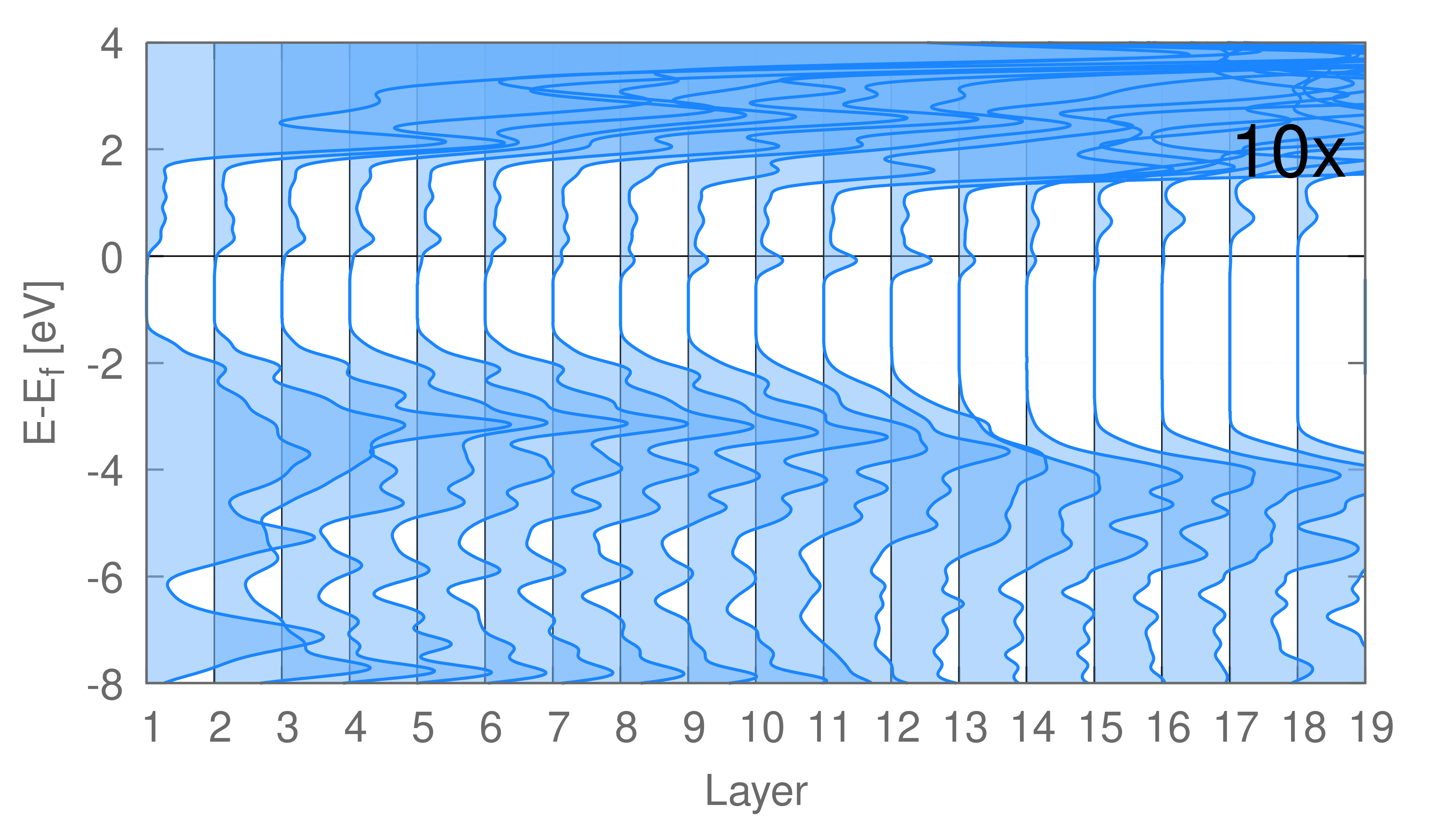}
\includegraphics[width=0.49\textwidth]{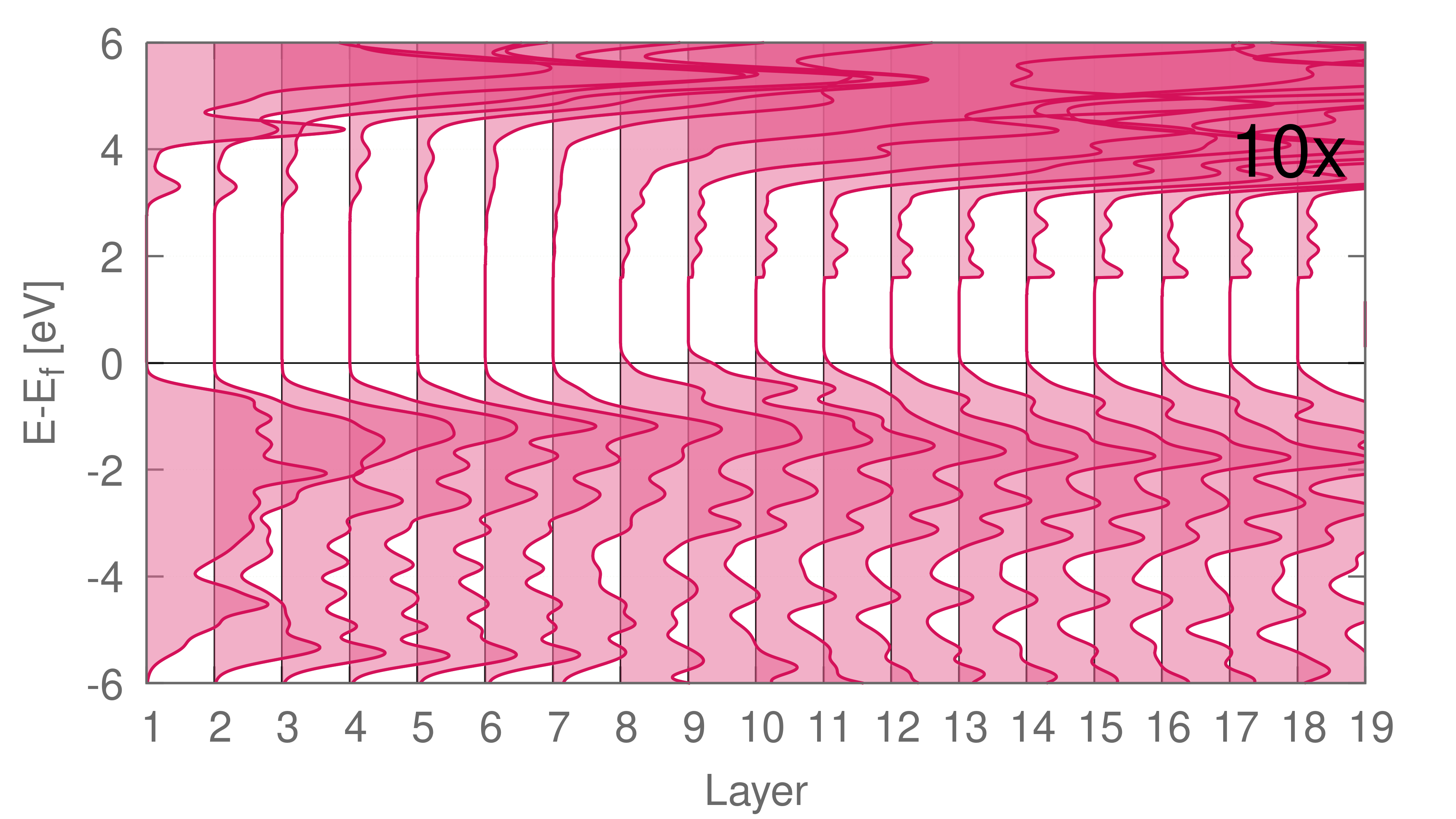} 
\caption{\label{fig:pdos} Total and atom-projected DOS (top) and layer-by-layer projected DOS (bottom) for the model slabs supporting a 2DEG (left) and a 2DHG (right). Here each ``layer'' implies the projection over both a Ga (or Al) plane and an adjacent N plane. For the first and last layers, projection over psH atoms is included. All curves have been smoothed using Gaussian broadening. In the bottom plots, conduction band DOS were enhanced by a factor 10.
}  
\end{figure}

We now turn to examining the electronic structure. 
Figure \ref{fig:pdos} (top) shows the total and atom-projected density of states (DOS) for the two systems considered.
The band gap is the same in both structures, and, noticeably, mid-gap states are not present.
Because of the fractional number of electrons considered, the Fermi level is located at the very top (bottom) of the valence (conduction) band, indicating the presence of free holes (electrons) in the system. 
To understand the real space location of the electronic states supporting the free carriers, Fig.\
 \ref{fig:pdos} (bottom) shows a layer-by-layer projection of the DOS, where here each ``layer'' implies the projection over both a Ga (or Al) plane and an adjacent N plane. For the first and last layers, projection over psH atoms is included. Clearly, the Fermi level resides within a band only near the interface, indicating the formation of a 2DCG, with no discernible bandgap states anywhere, including at the slab surfaces.

\begin{figure}
\includegraphics[width=0.49\textwidth]{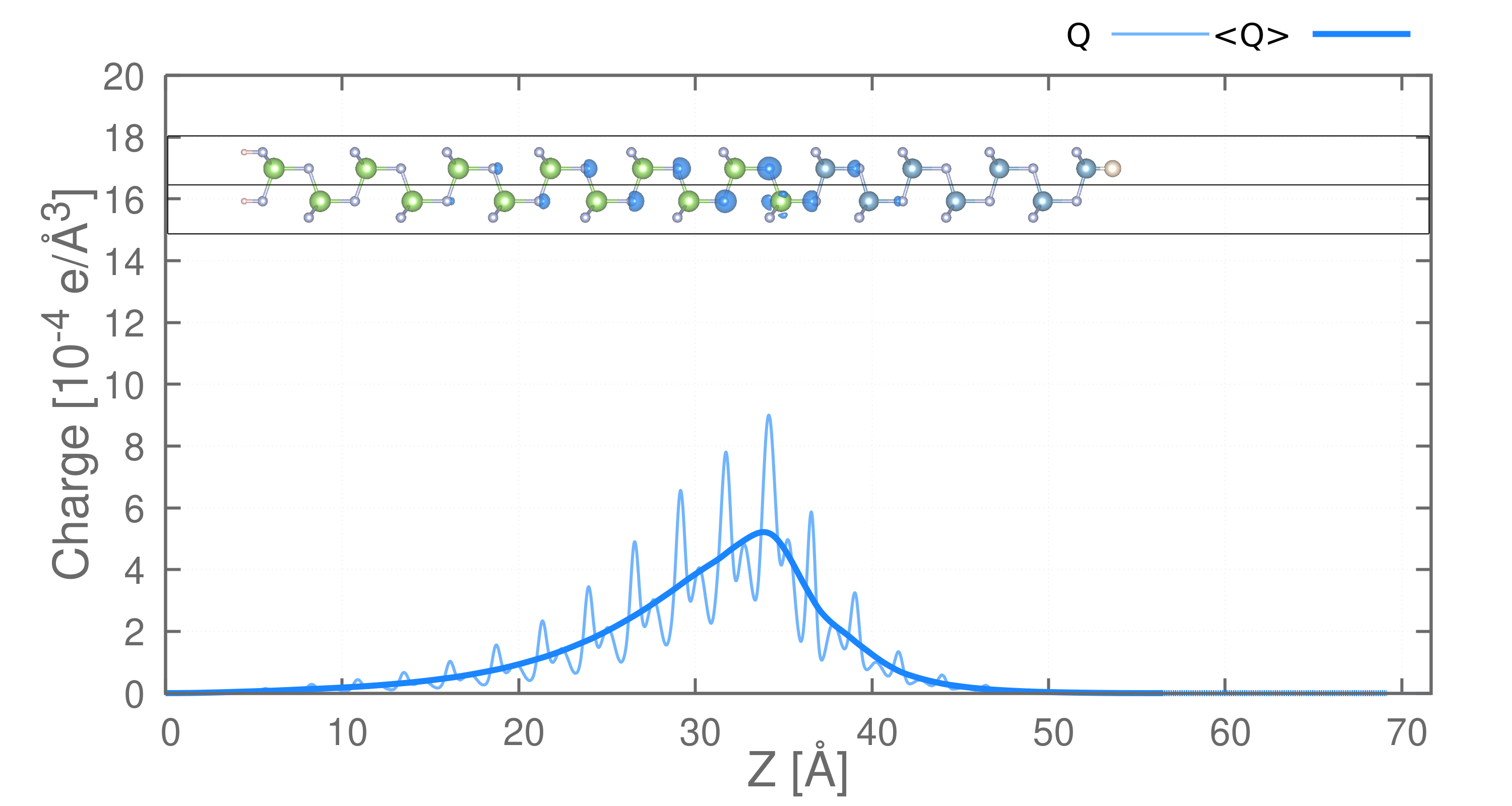} 
\includegraphics[width=0.49\textwidth]{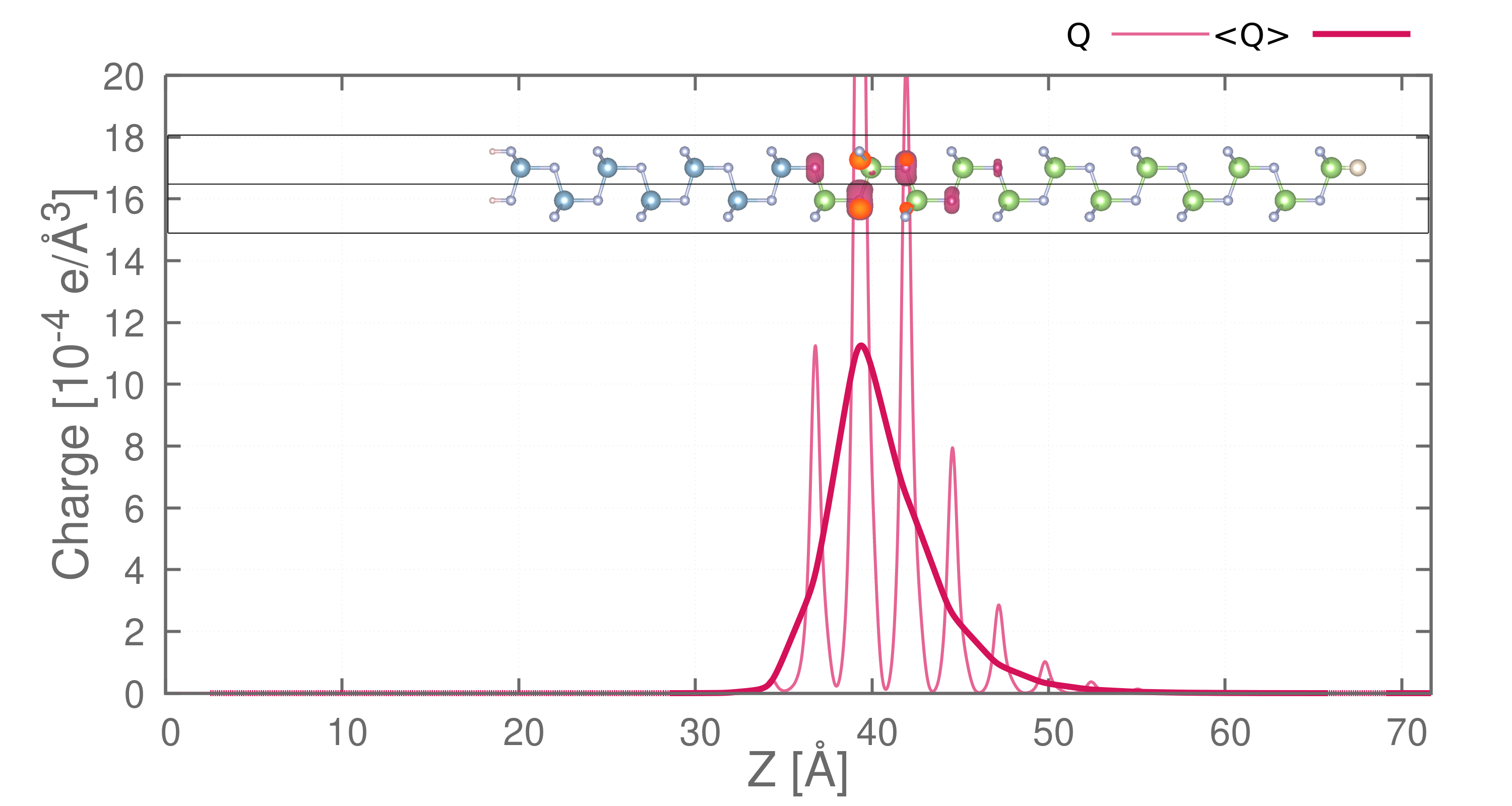} 
\caption{\label{fig:ed} Laterally-averaged local (thin line) and macroscopic (thick line) charge distribution of the 2DCG states along the z-axis. Overlaid figures report the three-dimensional distribution of the corresponding state. The right (left) panel can be identified with the 2DEG (2DHG) and represents occupied (empty) states of the conduction (valence) band. 
%
}
\end{figure}

To quantify the spatial extent of the 2DCG, we integrate the charge in a 25 meV window from the Fermi energy, with the results shown in Fig.\ \ref{fig:ed}. The free carriers accumulate at the interface and decay within a few atomic layers away from it. Electrons and holes accumulate for the and right  panel of the figure, respectively. 
The difference in localization is due to the different nature of the states where the carriers are hosted and has been previously pointed out~\cite{tonisch_polarization-induced_2012}.

The model employed here relies on the use of fictitious psH atoms, which still leaves the question of what would provide the interface charge in a real system. Clearly, charge must come from the external surfaces/interfaces, but it doesn't have to be supplied by in-gap surface states. For example, the charge could come from contact with a metal and/or from in-band resonance states,  or from chemical passivation, similar to what has been intensively discussed in the literature~\cite{harrison79theory,pashley1989electron,robertson2011bonding,robertson2015defect} for the zincblende GaAs case.
As outlined in Refs.~\cite{robertson2011bonding,robertson2015defect} such scenarios may also generate surface in-gap states, but their concentration can be minimized by various experimental procedures, and the emergence of a 2DCG does not depend on them.


To conclude, our first-principles calculations, based on a slab model passivated by fractional valence pseudo-hydrogen atoms, demonstrate that a high concentration of either the 2DEG or the 2DHG can be present at the interface of an undoped polar heterojunction at zero electric field, in the absence of bandgap surface states.
This result may help the modeling of electron devices based on the formation of 2DEG or 2DHG due to permanent polarization. 

\section*{Acknowledgements}

L.K.\ acknowledges support by the Aryeh and Mintzi Katzman Professorial chair and the Helen and Martin Kimmel Award for Innovative Investigation. A.M.R.\ acknowledges the support of the Army Research Laboratory via the Collaborative for Hierarchical Agile and Responsive Materials (CHARM) under cooperative agreement W911-NF-19-2-0119.
We thank Dr. C.E. Dreyer and Dr. S.-H. Yoo for  useful discussions.
The following article has been submitted to APL Materials.


\clearpage 
\bibliography{0paper2}

\end{document}